# Pedestrian Approach to the Two-Channel Kondo Model

P. Coleman[1], L. Ioffe[1] and A. M. Tsvelik[2]

[1] Serin Laboratory, Rutgers University, P.O. Box 849, Piscataway, New Jersey 08855-0849
[2] Department of Physics, University of Oxford, 1 Keble Road, Oxford OX1, UK

We reformulate the two-channel Kondo model to explicitly remove the unscattered charge degrees of freedom. This procedure permits us to move the non-Fermi liquid fixed point to infinite coupling where we can apply a perturbative strong-coupling expansion. The fixed point Hamiltonian involves a three-body Majorana zero mode whose scattering effects give rise to marginal self-energies. The compactified model is the $N = 3$ member of a family of "$O(N)$" Kondo models that can be solved by semiclassical methods in the large $N$ limit. For odd $N$, *fermionic* "Kink" fluctuations about the $N = \infty$ mean-field theory generate a fermionic $N$-body bound-state which asymptotically decouples at low energies. For $N = 3$, our semi-classical methods fully recover the non-Fermi liquid physics of the original two channel model. Using the same methods, we find that the corresponding $O(3)$ Kondo lattice model develops a spin-gap and a gapless band of coherently propagating three-body bound-states. Its strong-coupling limit offers a rather interesting realization of marginal Fermi liquid.

72.15.Nj, 71.30+h, 71.45.-d

## I. INTRODUCTION

The possible origins of non-Fermi liquid behavior in highly correlated metals is a topic of lively debate.[1-4] Non-Fermi liquid behavior has been observed in a variety of metallic systems, including low dimensional conductors,[5] cuprate superconductors,[6,7] systems at a quantum critical point[8] and in certain heavy fermion metals.[9,10] In this context, the two channel Kondo model is of particular interest,[11] for it provides a simple example of a class of non-Fermi liquid (NFL) behavior that is driven by local physics.[12-15] Variants of this model are of potential relevance to the physics of dilute heavy fermion systems[3,16,17] and two level tunneling systems.[18]

Most pedestrian techniques in condensed matter theory involve perturbative expansions about weak or strong coupling fixed points. Insight into the underlying physics is often gained when these simple approaches can be applied. Unfortunately, non-Fermi liquid behavior in the two-channel Kondo model is associated with an intermediate coupling fixed point,[11] placing it beyond the reach of elementary methods. In this paper show how this difficulty can be overcome by taking advantage of spin-charge decoupling. A key result is the derivation of an effective Hamiltonian for the low-energy spin physics of the two-channel Kondo model:

$$H^* = \sum_{\vec{k}} \epsilon_{\vec{k}} \vec{\Psi}_{\vec{k}} \cdot \vec{\Psi}_{\vec{k}} + \alpha \Phi \Psi^{(1)}(0) \Psi^{(2)}(0) \Psi^{(3)}(0). \quad (1)$$

Here $\vec{\Psi} \equiv (\Psi^{(1)}, \Psi^{(2)}, \Psi^{(3)})$ is a three-component Majorana fermion ($\Psi^{(a)} = \Psi^{(a)\dagger}$) with dispersion $\epsilon_{\vec{k}}$ and $\Phi = \Phi^\dagger$ is a Majorana fermion localized at the origin, whose singular scattering effects generate a marginal Fermi liquid. The spin density is given by $\vec{S}(x) = -i\vec{\Psi}(x) \times \vec{\Psi}(x)$.

We shall derive this Hamiltonian in two ways and show how it may be used to simply obtain the thermodynamics and spin-physics of the two-channel Kondo model.

Spin-charge decoupling, the separation of spin and charge into independent degrees of freedom at low energies, is a universal feature of impurity Kondo models. For multi-channel Kondo models, the use of a lattice cut-off interferes with this separation. By compactifying the two-channel model into a Hamiltonian that exclusively describes the decoupled spin degrees of freedom, we show how the over-screened fixed point[11] is eliminated, relocating the NFL fixed point to infinite coupling. We can then examine the NFL properties of the two channel Kondo model in a strong-coupling expansion. We can also apply semi-classical methods to examine the model both in weak coupling and in a lattice generalization.

Many aspects of the two channel Kondo model have been characterized by a combination of Bethe ansatz methods[12,13], conformal field theory[14], numerical diagonalization[22,23] and bosonization.[24,25] These methods show that the NFL ground state is characterized by

⋄ Logarithmic temperature dependence of the specific heat $C_v \propto T \ln T$ and magnetic susceptibility $\chi \propto \ln T$.

⋄ A remarkable zero point entropy of magnitude $\frac{1}{2} \ln 2$.

Emery and Kivelson's study[24] of a solvable anisotropic limit of the two-channel Kondo model identifies this fractional zero point entropy with a unique fermionic zero-mode: a "Majorana" fermion. This localized mode develops below the Kondo temperature ultimately decoupling from the conduction sea at the NFL fixed point. Sengupta and Georges[25] have shown that the residual low-energy coupling of this fermionic zero mode to the conduction sea creates the logarithmic temperature dependence of the susceptibility and specific heat. Our work complements these discoveries and provides a simple Hamiltonian formalism for studying them in the spin-isotropic limit.

We begin, in Section II by compactifying the two channel Kondo model, mapping the spin degrees of freedom of each channel onto the spin and isospin (charge) degrees of freedom of a single-flavored electron fluid. In section III we make a strong-coupling expansion of the non-Fermi fixed point, applying methods previously applied by Nozières to the one channel Kondo model. In the strong-coupling Hamiltonian, non-Fermi liquid behavior is associated with the development of a sharp *three-body* resonance in the conduction electron propagators. In Section IV we develop a versatile functional integral technique for studying this model in the regime of weak coupling. We first use a mean field approximation to find the apparent order parameter, then we show that $Z_2$ fluctuations of this order parameter restore the local symmetry and lead to the same non-Fermi liquid regime that was obtained in the strong coupling limit. Finally, in Section V we use the methods developed for an impurity model to formulate and solve its lattice generalization. Of particular interest here, is the competition between "incoherent" non-Fermi liquid behavior and coherent three-body bound-state motion. Section VI summarizes and concludes the results of these studies.

## II. MODEL

Our starting model is the two-channel Kondo model, which we write in a one-dimensional form

$$H' = it \sum_{n\lambda\sigma} [a^\dagger_{\lambda\sigma}(n+1) a_{\lambda\sigma}(n) - \text{H.c.}]$$
$$+ J[\vec{\sigma}_1(0) + \vec{\sigma}_2(0)] \cdot \vec{S}_d \quad (2)$$

where $\lambda = 1, 2$ labels two independent one dimensional conducting chains and $\vec{\sigma}_\lambda(0) = a^\dagger_{\lambda\alpha}(0)\vec{\sigma}_{\alpha\beta} a_{\lambda\beta}(0)$ is their spin density at the origin. In this model, prior to coupling to the local moment at the origin the spin degrees of freedom of the two chains are completely independent.

Motivated by the observation that in the continuum limit, the spin and charge degrees of freedom of a single chain behave as two decoupled spin-1/2 degrees of freedom, we now write down a compactified two-channel Kondo model where the local moment is coupled to the spin and isospin degrees of freedom of a single chain:

$$H = it \sum_n [c^\dagger_\sigma(n+1) c_\sigma(n) - \text{H.c.}]$$
$$+ J[\vec{\sigma}(0) + \vec{\tau}(0)] \cdot \vec{S}_d. \quad (3)$$

Here

$$\vec{\sigma}(x) = c^\dagger_\alpha(x)\vec{\sigma}_{\alpha\beta} c_\beta(x)$$

is the spin density at position $x$ and

$$\vec{\tau}(x) = \tilde{c}^\dagger_\alpha(x)\vec{\sigma}_{\alpha\beta}\tilde{c}_\beta(x)$$

is the isospin density, whose diagonal and off-diagonal components respectively describe the charge and pair density at site $x$. The tilde on the electron operators denotes a Nambu spinor

$$\tilde{c} = \begin{pmatrix} c_\uparrow \\ c^\dagger_\downarrow \end{pmatrix}. \quad (4)$$

We link the two models by making an identification of the low-energy spin modes in (2) with the low energy spin and isospin modes of (3).

$$\begin{aligned} \vec{\sigma}_1(x) &\longleftrightarrow \vec{\sigma}(x) \\ \vec{\sigma}_2(x) &\longleftrightarrow \vec{\tau}(x) \end{aligned} \quad (5)$$

The dynamical equivalence of these two sets of operators is established by comparing their long-distance, long-time correlators at $J = 0$. At $J = 0$, in a particle-hole symmetric band, the correlators of isospin and spin are identical. Furthermore, since the low energy spin and charge degrees of freedom are decoupled, the joint low energy correlators of isospin $\vec{\tau}(0)$ and spin $\vec{\sigma}(0)$ completely factorize, thereby emulating the joint correlations of the spin density in the original two channel Kondo model. More precisely,

$$\langle \vec{\sigma}(1)\vec{\sigma}(2)\ldots\vec{\sigma}(n)\vec{\tau}(1')\vec{\tau}(2')\ldots\vec{\tau}(r')\rangle_{J=0}$$
$$= \langle \vec{\sigma}_1(1)\vec{\sigma}_1(2)\ldots\vec{\sigma}_1(n)\rangle\langle \vec{\sigma}_2(1')\vec{\sigma}_2(2')\ldots\vec{\sigma}(r')\rangle_{J=0} \quad (6)$$

if all time scales are longer than the band-width (i.e. all $|\tau_i - \tau_k| \gg 1/t$). (See appendix for a more detailed discussion). If we now turn on the interaction, this one-to-one correspondence between the correlation functions ensures that the models are equivalent in the continuum limit, to all orders in perturbation theory. By explicitly separating the spin degrees of freedom in the compactified model, when we now use a lattice cut-off procedure we avoid developing an unrenormalizable coupling between spin and charge in the strong-coupling limit.

It is convenient to rewrite the conduction electron spinor in terms of its real and imaginary components $\Psi$ as follows

$$c(n) = \frac{1}{\sqrt{2}}\left(\Psi^0(n) + i\vec{\Psi}(n)\cdot\vec{\sigma}\right)\begin{pmatrix} 0 \\ -i \end{pmatrix} \quad (7)$$

In terms of these components, (see Appendix B)

$$\vec{\sigma}(n) + \vec{\tau}(n) = -i\vec{\Psi}(n) \times \vec{\Psi}(n), \quad (8)$$

and

$$(c^\dagger_\sigma(n+1) c_\sigma(n) - \text{H.c.}) = \sum_{a=0}^{a=3} \Psi^a(n+1)\Psi^a(n). \quad (9)$$

The Hamiltonian is seen to decouple into a "scalar" and "vector" part

$$H = H_{sc} + H_v. \qquad (10)$$

where

$$H_{sc} = it \sum_n \Psi^0(n+1)\Psi^0(n), \qquad (11)$$

$$H_v = it \sum_n \vec{\Psi}(n+1) \cdot \vec{\Psi}(n) - iJ\vec{S}_d \cdot \left(\vec{\Psi}(0) \times \vec{\Psi}(0)\right). \qquad (12)$$

The "scalar" part (11) describes a freely propagating fermion mode whereas the "vector" part (12) describes a local moment interacting with a sea of vector Majorana fermions. This simple result shows that the low energy spin excitations in the the two channels fuse to form a band of free singlet and interacting vector fermions. Schematically:

$$\text{spinon } [1] \otimes \text{spinon } [2] \longrightarrow \text{scalar} \oplus \text{vector}. \qquad (13)$$

From (8), we see that the joint spin-density of the two chains is carried by these vector excitations.

The universal scaling properties of this model are identical to those of the two channel Kondo model. The weak coupling scaling equation for the interaction constant $g \equiv \rho J$ (where $\rho$ is the density of conduction electron states at the Fermi energy) has a form

$$\frac{\partial g(\xi)}{\partial \xi} = \beta(g) = 2g^2 - g^3 + O\left[g^4\right]. \qquad (14)$$

On general grounds, the first two terms in the weak coupling expansion of the beta function are universal.[19] Higher order terms which determine the location of the NFL fixed point can not be defined in a *universal* way, so the precise location of the NFL fixed point depends on the cut-off procedure. The original and compactified models correspond to different regularization schemes with running coupling constants $g_1(\xi)$ and $g_2(\xi)$ that differ in strong-coupling due to the higher order terms in the beta functions. Their coupling constants coincide at weak coupling,

$$g_2 = G[g_1], \qquad (15)$$
$$G[x] = x + O[x^3], \qquad (16)$$

but the higher order terms of $G[x]$ depend on the cut-off procedure. In the compactified model there is no unstable "over-screened" fixed point. Though the long-time correlation functions of $\vec{S} = \frac{1}{2}[\vec{\sigma}(0) + \vec{\tau}(0)]$ mimic the correlations of the operator $\vec{S}' = \frac{1}{2}[\vec{\sigma}_1(0) + \vec{\sigma}_2(0)]$, in the two-channel problem $S$ is a spin-1/2 operator obeying usual identities:

$$[S^a, S^b] = i\epsilon_{abc}S^c, \qquad (17)$$
$$S^2 = \frac{3}{4}. \qquad (18)$$

On a lattice, the spin and isospin can not simultaneously be finite at one site, so it is not possible to produce a spin 1 composite at the origin to over-screen the local moment. Consequently, the unstable "over-screened" fixed point vanishes from the renormalization flows. Instead, as we shall show, the non-Fermi liquid behavior appears as a stable fixed point at infinite coupling $g_2 = \infty$. Qualitatively, if $g_1^*$ is the zero of the beta function for the original two-channel Kondo model, by changing the cut-off procedure we have mapped this point to infinity $g_2^* = G[g_1^*] = \infty$.

### III. STRONG COUPLING LIMIT

In this section we examine the strong coupling limit of the compactified two-channel Kondo model. For simplicity we assume that the local moment is located at the origin of a chain:

$$H = it \sum_{n=0}^{n=\infty} \vec{\Psi}(n+1) \cdot \vec{\Psi}(n) - iJ\vec{S}_d \cdot \left(\vec{\Psi}(0) \times \vec{\Psi}(0)\right). \qquad (19)$$

We imagine that we carry out a Wilson scaling procedure,[19-21] keeping track of the corrections to the free energy and the evolution of the Hamiltonian as we integrate out the high energy degrees of freedom and move away from weak coupling. On general grounds, the qualitative physics described by the Hamiltonian at each point on the scaling trajectory will not change. If the strong-coupling fixed point is stable we can obtain a simple description of the essential physics of the model at weak coupling.

We begin by examining the character of the strong coupling fixed point. At $J = \infty$ we may neglect all sites except $n = 0$:

$$H_\infty = -iJ\vec{S}_d \cdot (\vec{\Psi} \times \vec{\Psi}) \qquad (20)$$

The eigenstates of the strong coupling Hamiltonian $H_{str}$ correspond to singlet and triplet states formed between the local moment and either the spin or isospin of conduction electron at the origin. Thus, there are two singlet and two triplet states, separated by energy $2J$, as shown in Fig. 1. This two-fold duplication of energy levels results from a local supersymmetry: there are two real Fermi operators, $\Psi^{(0)}$, and

$$\Phi = -2i\Psi^{(1)}\Psi^{(2)}\Psi^{(3)},$$

which commute with the Hamiltonian $H_\infty$ (20). The *complex* fermion $\zeta = (\Psi^{(0)} - i\Phi)/\sqrt{2}$ transforms one state into its degenerate partner. Unlike bosonic degeneracies, transitions between the two singlet states is always accompanied by emission/absorption of an *odd* number of fermions.

To examine the effect of hybridization we first reinstate it between sites 0 and 1, writing

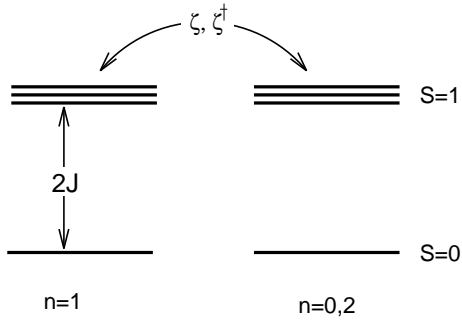

Fig. 1.: **Energy levels for the strong coupling limit.** The singlet can form with a spin ($n = 1$) or "isospin" ($n = 0, 2$). The two-fold degeneracy can be identified with the fermion $\zeta$ which commutes with the strong coupling Hamiltonian.

$$H = H_\infty + \hat{t}, \qquad (21)$$

where $\hat{t} = it\vec{\Psi}(1) \cdot \vec{\Psi}(0)$. The hopping term $\hat{t}$ causes transitions amongst the singlet and triplet states at the origin. Its matrix element between the singlet states is zero. We now use perturbation theory in $t/J$ to eliminate the high energy states triplet states and form a low-energy effective Hamiltonian $H^*$. This is done by making the canonical transformation: $H \rightarrow H^* = e^S H e^{-S}$, where

$$S = \hat{P}_o \left[ \left(\frac{\hat{t}}{-2J}\right) + \left(\frac{\hat{t}}{-2J}\right)^2 + \ldots \right] \hat{P}_1 - (\text{H.c.}) \qquad (22)$$

is chosen to eliminate the off-diagonal terms between the singlet and triplet states. Here $\hat{P}_1$ and $\hat{P}_0$ respectively project into the subspaces with a triplet, or singlet configuration at the origin. The residual interaction induced in the singlet subspace is then

$$H_{int} = \hat{P}_0 \left[ -\frac{\hat{t}^2}{2J} + \frac{\hat{t}^3}{(2J)^2} + \ldots \right] \hat{P}_0. \qquad (23)$$

Under the singlet projection, the first term in (23) vanishes. The second term becomes

$$H_{int} = \alpha \Phi \Psi^{(1)}(1) \Psi^{(2)}(1) \Psi^{(3)}(1), \qquad (24)$$

where $\alpha = 3t^3/4J^2$. This term couples to the fermionic zero mode $\Phi$ to a three-body composite of conduction electrons at site 1. In the more conventional one channel Kondo model, this term is absent and the leading term in the Hamiltonian is a benign Fermi liquid interaction proportional to $t^4/J^3$. The three-body resonance feature of the two-channel fixed point fundamentally transforms its physics. Our final fixed point Hamiltonian for the spin-degrees of freedom of the two-channel Kondo model is thus

$$\begin{aligned} H^* = it \sum_{n=1}^{n=\infty} \vec{\Psi}(n+1) \cdot \vec{\Psi}(n) \\ + \alpha \Phi \Psi^{(1)}(1) \Psi^{(2)}(1) \Psi^{(3)}(1), \end{aligned} \qquad (25)$$

where the site at the origin is explicitly excluded.

Let us now consider the perturbative corrections that result from coupling the itinerant band to the zero mode. We shall show that these corrections are formally 0 in the renormalization group sense, but that they produce strong scattering effects on the conduction electrons. We construct the perturbation theory using Feynman diagrams containing the local propagators of the itinerant conduction electrons and the three-body bound state. Both propagators describe Majorana fermions, and are represented by lines without arrows. We represent the conduction electron local propagator is $\langle \Psi^a(1,\tau) \Psi^b(1,0) \rangle = \delta_{ab} G(\tau)$ as

$$(1,0) \xrightarrow{\quad b \quad\quad a \quad} (1,\tau) \; = \; \delta_{ab} G(\tau), \qquad (26)$$

where

$$G(\tau) = \int_{-\pi}^{\pi} \frac{dk}{2\pi} [1 - f(\epsilon_{\vec{k}})] e^{-\epsilon_{\vec{k}}\tau} = \left[\frac{\rho \pi T}{\sin(\tau \pi T)}\right]. \qquad (27)$$

Here T is the temperature, $\epsilon_{\vec{k}} = -2t \sin k$ and $\rho = \frac{1}{2\pi v_F}$ is the density of states for a Fermi velocity $v_F = 2t$. The bare three-body bound-state propagator $G_\Phi^0(\tau) = \langle \Phi(\tau) \Phi(0) \rangle_0$ is

$$0 \;-\;-\;-\;-\;-\; \tau \; = \; G_\Phi^0(\tau), \qquad (28)$$

where

$$G_\Phi^0(\tau) = -T \sum_n \frac{1}{i\omega_n} e^{-i\omega_n \tau} = \frac{1}{2} \text{sgn}(\tau), \qquad (29)$$

At T $= 0$, $G(\tau) \sim \frac{1}{\tau}$, so the scaling dimension of the conduction electron operator $\Psi(1)$ is $d_\Psi = \frac{1}{2}$, whilst the fermionic zero mode is dimensionless: $d_\Phi = 0$. The dimension of the interaction term in the action $S_{int} \propto \int d\tau \Phi \Psi^1 \Psi^2 \Psi^3$ is consequently

$$d_{int} = 1 - d_\Phi - 3d_\Psi = -1/2.$$

The interaction is formally "irrelevant", because it scales as $[\tau^{-1/2}] \equiv [\omega^{1/2}]$. From a practical standpoint, this means that each additional power of $\alpha^2$ associated with vertex corrections to the skeleton diagrams introduces an additional power of frequency into the diagram. The vertex corrections to the bare interaction vertex between three conduction electrons and the zero mode, shown in Fig. 2 are thus proportional to $\alpha^2 \omega$ and can be completely neglected at low energies.

The polarizability of the electron fluid leads to two leading corrections to the interactions between the localized and itinerant fermions, as shown in Fig. 3. The vertex induced between the localized Majorana modes and the itinerant electrons (shown in Fig. 3a) is $\alpha^2 \chi_{ee}(\tau)$, where $\chi_{ee}(\tau)$ is the local spin-susceptibility of the conduction electrons. At long times, $\chi_{ee}(\tau)$ becomes small: $\chi_{ee}(\tau) \propto 1/\tau^2$ which implies a frequency dependence $\chi_{ee}(\nu_n) \propto |\nu_n|$ showing that this term is irrelevant at low frequencies. The second vertex shown in Fig. 3b depends on the joint-susceptibility between the conduction electrons and the three-body bound-state $\chi_{e\Phi}$. From a scaling argument, $\chi_{e\Phi}(\tau) \propto 1/|\tau|$ which implies an interaction $g^2\chi_{e\Phi}(\omega) \propto \ln|\omega|$. This term generates a marginal interaction between the conduction electrons that is responsible for departures from Fermi liquid behavior.

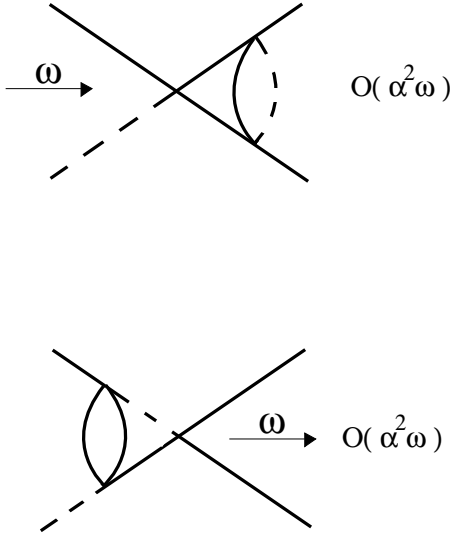

Fig. 2.: **Vertex corrections to the bare interaction.** Solid lines denote the electron propagator $G(\tau)$. Dashed lines denote the three-body propagator $G^o_\Phi(\tau) = \tfrac{1}{2}\mathrm{sgn}(\tau)$. The vertex corrections scale as $O(\alpha^2\omega)$.

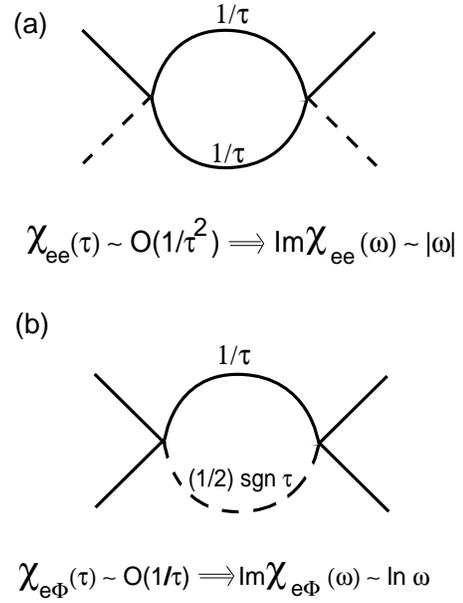

Fig. 3.: **Renormalization of interactions** (a) Irrelevant susceptibility $\chi_{ee}(\tau) \propto \tfrac{1}{\tau^2}$; (b) marginal susceptibility between conduction electron and three-body bound state $\chi_{e\Phi}(\tau) \propto \tfrac{1}{\tau}$.

Another way to consider the effect of these interactions is to evaluate the self energy parts that they generate. Only two diagrams are possible in the second order (Fig. 4). The term shown in Fig. 4a renormalizes the propagator of the fermionic zero mode.

$$G^0_\Phi(\omega) \rightarrow G_\Phi(\omega) = \frac{1}{\omega - \Sigma_\Phi(\omega)}, \qquad (30)$$

Since $\Sigma_\Phi(\tau) \propto 1/\tau^3$, $Im\Sigma_\Phi(\omega) \propto \omega^2$ at low frequencies. This means that the sharp pole in the zero-mode propagator is preserved to all orders in the scattering processes

$$\frac{1}{\pi}\mathrm{Im}\big[G_\Phi(\omega)\big] = Z\delta(\omega) + (\text{background}) \qquad (31)$$

where $Z^{-1} = [1 - \delta_\omega \Sigma_\Phi(\omega)]|_0$. In other words, a renormalized component of the three-body operator *commutes* with the fixed point Hamiltonian. By contrast $\Sigma_\Psi(\tau) \propto 1/\tau^2$, and hence $Im\Sigma_\Psi(\omega) \propto |\omega|$,

$$\Sigma_\Psi(\omega) \propto \omega\ln\omega, \qquad (32)$$

a signature of a marginal Fermi liquid. The stability of the fermionic zero mode guarantees that this feature is not removed by higher order processes. A consequence is that the itinerant spinons described by the conduction sea develop a sharp, zero energy pole in their *three-body* propagator.

The explicit form of the off-diagonal susceptibility $\chi_{e\Phi}$ in the frequency domain is of particular interest. We write

$$\chi_{e\Phi}(\tau)\delta^{ab} = \langle \big[\Psi^a(1,\tau)\Phi(\tau)\big]\big[\Psi^b(1,0)\Phi(0)\big]\rangle, \qquad (33)$$

so that

$$\chi_{e\Phi}(\nu_n) = -\int_0^\beta d\tau G(\tau)G^0_\Phi(\tau)e^{i\nu_n\tau}. \qquad (34)$$

Writing this as a Matsubara sum

$$\chi_{e\Phi}(\nu_n) = \rho T \sum_{|\omega_m|<D} \frac{\mathrm{sgn}(\omega_m)}{\omega_m + \nu_n}$$
$$= \frac{\rho}{\pi}\left[\ln\frac{D}{2\pi T} - \Psi\Big(\frac{1}{2} + \frac{|\nu_n|}{2\pi T}\Big)\right], \qquad (35)$$

where $\Psi(x)$ denotes the digamma function.[26] Making the analytic continuation gives

$$\chi_{e\Phi}(\nu - i\delta) = \frac{\rho}{\pi}\left[\ln\frac{D}{2\pi T} - \Psi\Big(\frac{1}{2} - \frac{i\nu}{2\pi T}\Big)\right]. \qquad (36)$$

The imaginary part of this susceptibility is

$$\chi''_{e\Phi}(\nu) = \rho\left[\frac{1}{2}\tanh\Big(\frac{\nu}{T}\Big)\right], \qquad (37)$$

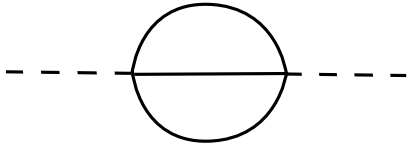

(a) $\Sigma(\tau) \sim O(1/\tau^3) \Longrightarrow Im\Sigma(\omega) \sim \omega^2$

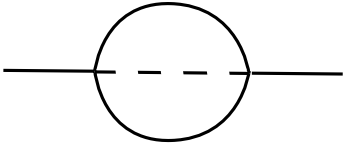

(b) $\Sigma(\tau) \sim O(1/\tau^2) \Longrightarrow \Sigma(\omega) \sim \omega \ln \omega$

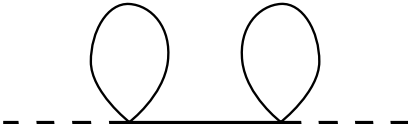

(c) $\Sigma(\omega_n) = -\imath \Delta \, sgn(\omega_n)$

Fig. 4.: **Self energy corrections** (a) Irrelevant correction to three-body self energy. Since $Im\Sigma(\omega) \sim \omega^2$, this term preserves the sharp zero-energy pole in the three-body propagator. (b) Marginal self energy of conduction electrons induced by scattering off sharp three-body bound-state. $Im\Sigma(\omega) \sim |\omega|$. (c) "Hartree" self energy that gives Majorana bound-state a finite width $\Delta \propto B^2$ in a magnetic field $B$.

i.e.

$$\chi''_{e\Phi}(\nu) \sim \begin{cases} \frac{\nu}{T}, & (\nu << T) \\ \text{constant}. & (\nu >> T) \end{cases} \quad (38)$$

The "marginal" form$^2$ of this susceptibility is a direct consequence of scattering off the perfectly sharp three-body bound-state.

Suppose we introduce a magnetic field by adding the term of a magnetic field that couples to both conduction electrons and the local moment,

$$H_B = \vec{B} \cdot [\vec{S}_d - (i/2) \sum_{j=0}^{\infty} \vec{\Psi}_j \times \vec{\Psi}_j]. \quad (39)$$

The splitting between the singlet and triplet states at site 0 is unaltered up to terms of order $O(B^2/J)$, so the residual coupling of the magnetic field in the fixed point Hamiltonian only involves the electrons at sites $x > 0$

$$H_B^* = -\frac{i}{2}\vec{B} \cdot \sum_{n=1}^{\infty} \vec{\Psi}(n) \times \vec{\Psi}(n) + O(B^2/J). \quad (40)$$

These terms introduce an off-diagonal contribution to the conduction electron propagators

$$\langle \psi_a(1,\tau)\psi_b(1,0)\rangle = G_o(\tau)\delta_{ab} + i\epsilon_{abc}B_c\rho e^{-D|\tau|}\text{sgn}\tau, \quad (41)$$

where $\rho = 1/(2\pi v_F)$ is the density of states and $D$ the bandwidth. The second term in the propagator gives rise to a static magnetization

$$-\frac{i}{2}\langle \vec{\psi}(1) \times \vec{\psi}(1)\rangle = \vec{B}\rho, \quad (42)$$

and the zero-mode self-energy now acquires a "Hartree" self energy, as shown in Fig. 4. This term breaks the fermionic degeneracy and broadens the sharp resonance into a Lorentzian of width $\Delta = \pi B^2(\alpha\rho)^2\rho$. The field-dependent energy scale will act as an effective Fermi temperature

$$T_F \sim \Delta(B) \propto B^2. \quad (43)$$

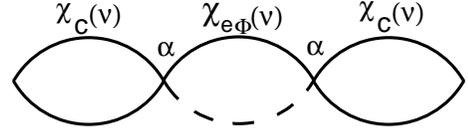

$\chi_{spin}(\nu) = \alpha^2 \chi_c^2(\nu) \chi_{e\Phi}(\nu)$

Fig. 5.: **Impurity susceptibility.** Showing how the magnetic field couples to the marginal susceptibility.

One of the most interesting physical quantities to calculate is the dynamical magnetic susceptibility of the impurity. This is given by the diagram shown in Fig. 5, which leads to

$$\chi_{spin} = \alpha^2 \chi_c^2(\nu)\chi_{e\Phi}(\nu), \quad (44)$$

where $\chi_c(\nu)$ is the uniform magnetic susceptibility of the conduction band. At energies $|\nu| << t$, $\chi_c(\nu) = \rho$, so that the impurity spin susceptibility is directly proportional to the marginal susceptibility $\chi_{e\Phi}(\nu)$

$$\chi_{spin}(\nu) = (\alpha\rho)^2\chi_{e\Phi}(\nu). \quad (45)$$

From this result, we may also deduce the the static susceptibility

$$\chi_{spin} = \rho(\alpha\rho)^2\ln\left[\frac{\Lambda}{T}\right], \quad (46)$$

where $\Lambda = D/(2\pi e^{\Psi(\frac{1}{2})})$.

We may calculate the dominant non-Fermi liquid corrections to the free energy within second order perturbation theory. These calculations essentially duplicate the earlier work by Sengupta and Georges.[25] In a weak magnetic field $B << T$, there are two leading contributions to the free energy ( Fig. 6.), a zero field part (Fig. 6(a))

$$F_1(T) = \frac{1}{2}\alpha^2 \int_0^\beta [G(\tau)]^3 [\frac{1}{2}\mathrm{sgn}\tau]d\tau$$
$$= \left[\frac{\rho(\alpha\rho\pi\mathrm{T})^2}{2}\right] \int_{T/\Lambda}^{\pi/2} dx \frac{1}{\sin^3 x}$$
$$= -\frac{\rho(\alpha\pi\rho\mathrm{T})^2}{4} ln\left[\frac{\Lambda}{\mathrm{T}}\right], \qquad (47)$$

and a finite field part (Fig. 6(b))

$$F_2(T) = \frac{(\alpha M)^2}{2}\int_0^\beta G(\tau)[\frac{1}{2}\mathrm{sgn}\tau]d\tau \qquad (M = \rho B)$$
$$= \left[\frac{\rho(\alpha\rho B)^2}{2}\right]\int_{T/\Lambda}^{\pi/2} dx \frac{1}{\sin x}$$
$$= -\frac{\rho(\alpha\rho B)^2}{2} ln\left[\frac{\Lambda}{\mathrm{T}}\right]. \qquad (48)$$

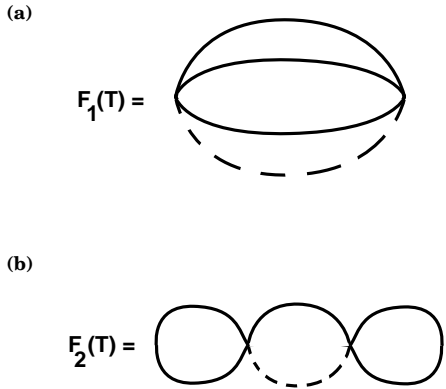

Fig. 6.: **Leading dangerous correction to free energy.** (a) Zero field component: leading correction from scattering off the three-body bound-state; (b) Finite field component.

Adding these two contributions to the free energy of the asymptotically decoupled zero mode, $F_0 = E_o - \frac{T}{2}\ln T$ gives

$$F = E_o - \frac{T}{2}\ln T - \frac{\rho(\alpha\rho)^2}{4}\left[(\pi T)^2 + 2B^2\right] ln\left[\frac{\Lambda}{\mathrm{T}}\right] \qquad (49)$$
$$(B << T)$$

This leads to logarithmic contributions to the susceptibility and linear specific heat $(\gamma(T) = C_V(T)/T)$

$$\begin{pmatrix} \chi \\ \gamma(T) \end{pmatrix} = \begin{pmatrix} 2 \\ \pi^2 \end{pmatrix} \frac{\rho}{2}(\rho\alpha)^2 ln\left(\frac{\Lambda}{\mathrm{T}}\right) \qquad (50)$$

The bulk susceptibility and linear specific heat of the original two channel Kondo model are

$$\chi_o = 2N\rho$$
$$\gamma_o = 8N\left(\frac{\pi^2}{3}\right)\rho \qquad (51)$$

where $N$ is the number of sites in the lattice, giving a Wilson ratio

$$W = \frac{\chi_i/\chi_o}{\gamma_i/\gamma_o} = \left(\frac{2}{\pi^2}\right)\left(\frac{2\pi}{3}\right)^2 = \frac{8}{3} \qquad (52)$$

At fields large compared with the temperature, the Majorana zero mode behaves as a resonant level model with a finite width $\Delta \propto B^2$. In this case, the repeated scattering off the resonance must be summed to all orders. This gives rise to a resonant-level contribution to the free energy

$$F[\Delta] = -\frac{T}{2}\sum_{-\infty}^\infty ln\left[(i\omega_n + i\Delta_n)\right] \qquad (53)$$

where $\Delta_n = \mathrm{sgn}(\omega_n)\theta(D - |\omega_n|)\Delta$. Using the result

$$ln\left[\frac{\Gamma(z+k)}{\Gamma(z)}\right] = \sum_{0 \le n < k} ln[z+n] \qquad (54)$$

we can carry out the Matsubara sum to obtain

$$F[\Delta] = \mathrm{T}ln\left\{\frac{\Gamma(\frac{1}{2} + \frac{\Delta\beta}{2\pi})}{\sqrt{2\pi}}\right\} - \frac{\Delta}{2\pi}ln\left(\frac{D}{2\pi T}\right), \qquad (55)$$

provided $(\Delta(B) << T)$. The function

$$S = -\frac{\partial F}{\partial T} = -ln\frac{\Gamma(\frac{1}{2} + \frac{\Delta\beta}{2\pi})}{\sqrt{2\pi}} + \frac{\Delta\beta}{2\pi}\left[\psi(\frac{1}{2} + \frac{\Delta\beta}{2\pi}) - 1\right] \qquad (56)$$

describes the field-induced quenching of the $\frac{1}{2}ln 2$ entropy associated with the three-body bound-state.

To complete our discussion of the non-Fermi liquid fixed point, we should like to consider the entropy associated with the fermionic mode. In the single site ($J = \infty$) limit there are two real fermions, $\Psi^{(0)}$ and $\Phi$ that commute with the Hamiltonian, which gives an entropy $ln 2$. Once the coupling to the remainder of conduction sea is included, the scalar fermion $\Psi^{(0)}$ becomes a part of the decoupled Fermi band, so half the fermionic degeneracy is removed. To further elucidate the nature of the "fractional" degeneracy it is instructive to consider two "two channel" Kondo impurities located at opposite ends of the conduction chain (Fig. 7). In such a situation there are two fermionic zero modes $\Phi(0)$ and $\Phi(N)$ associated with each end of the chain. Using perturbation theory we may deduce that these two modes are coupled by

$$H_\Phi = -i\lambda\Phi(0)\Phi(N), \qquad (57)$$

where $\lambda = \alpha \int G^3(\tau) d\tau$ and

$$G(\tau) = \langle \Psi^{(1)}(N-1,\tau)\Psi^{(1)}(1,0)\rangle$$
$$= -\frac{2i}{\pi}\left[\frac{N}{(v_F\tau)^2 + N^2}\right] + O(1/N^2). \quad (58)$$

Thus $\lambda \sim \alpha^2/(v_F N^2)$. The interaction (57) leads to a splitting of two zero modes. To show this, it is convenient to form a complex fermion $\xi = \frac{1}{\sqrt{2}}[\Phi(0) - i\Phi(N)]$, then

$$H_\Phi = \lambda(\xi^\dagger \xi - \frac{1}{2}), \quad (59)$$

indicating that as $N \to \infty$ the ground state of this system will develop a two fold degeneracy and, hence, a zero point entropy $S = \ln 2$. This entropy is completely delocalized, and for this reason we must associate a $\frac{1}{2}\ln 2$ entropy with the residual real Majorana zero mode $\tilde\Phi(0)$. The reasoning employed here is strongly reminiscent of arguments used to assign a fractional charge to a soliton in a polyacetylene chain.[27]

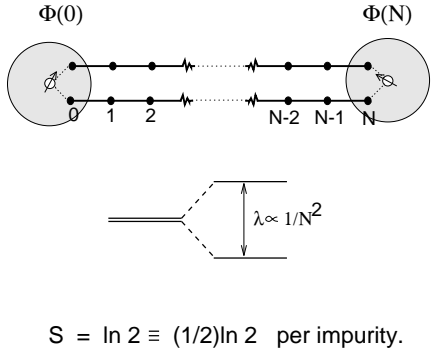

Fig. 7.: **Fractional entropy.** A two channel Kondo chain with $N+1$ sites. For finite chain length the two Majorana zero-modes are coupled, giving rise to a splitting of the ground state of order $O(1/N^2)$. In the limit of $N \to \infty$, every state in the Hilbert space is two-fold degenerate, leading to a "delocalized entropy" of $\frac{1}{2}\ln 2$ per local moment.

## IV. WEAK COUPLING LIMIT

In this Section we solve a generalization of the spin-charge Kondo model (12) in which the number of components in the vector Majorana is generalized from three to $N$. We then formulate the model as a path integral and show that the fluctuations about the large $N$ limit (i.e. about the saddle point solution) recover the key aspect of the strong coupling limit, i.e. they reproduce the effective low energy Hamiltonian (25). In contrast to the previous section, this method can be applied to the weak coupling properties of the model.

To develop this generalization, we introduce a Majorana representation of the local moments[34]

$$\vec{S}_d = -\frac{i}{2}\vec\eta \times \vec\eta, \quad (60)$$

where the operators $\vec\eta$ are real and satisfy canonical anti-commutation rules $\{\eta^a, \eta^b\} = \delta_{ab}$. In terms of these operators the Hamiltonian can be written

$$H = it\sum_n\left[\vec\psi(n+1)\cdot\vec\psi(n)\right] - J(i\vec\eta\cdot\vec\Psi(0))^2. \quad (61)$$

We now generalize the number of components of $\eta^a$ and $\Psi^a$ from 3 to $N$: $a = 1\ldots N$. Then the generalized Hamiltonian

$$H = it\sum_n\left[\vec\psi(n+1)\cdot\vec\psi(n)\right] - \frac{\tilde{J}}{2N}(i\vec\eta\cdot\vec\Psi(0))^2, \quad (62)$$

has a global $O(N)$ invariance

$$\eta^a \to R_{ab}\eta^b, \quad (63)$$
$$\Psi^b \to R_{ab}\Psi^b, \quad (64)$$

where $R_{ab}$ is an $N$ dimensional orthogonal matrix. We have rescaled $J \to \tilde{J}/2N$ so that with $\tilde{J}$ fixed, the Hamiltonian grows extensively with $N$, establishing a well defined mean theory in the large $N$ limit.

The interaction term can be conveniently rewritten in terms of $O(N)$ spins

$$S_d^{ab} = -i[\eta^a, \eta^b], \quad (65)$$
$$S_c^{a,b}(j) = -i[\Psi^a(j), \Psi^b(j)], \quad (66)$$

whereupon

$$H_{int} = \frac{\tilde{J}}{2N}S_d^{ab}S_c^{ab}(0). \quad (67)$$

The $O(N)$ Kondo model possesses a local $Z_2$ invariance under transformation

$$\vec\eta \to \hat{P}\vec\eta\hat{P}^{-1} = -\vec\eta. \quad (68)$$

The explicit form of the $Z_2$ operator depends on whether $N$ is even or odd. For the case of $N$ even

$$\hat{P} = 2^{N/2}\eta^1\eta^2\ldots\eta^N. \quad (69)$$

The case of odd $N$ requires special consideration. To construct the Hilbert space which represents the Majorana fermions $\eta^1, \eta^2 \ldots \eta^N$, we may pair $N-1$ fermion $\eta^2 \ldots \eta^N$ into complex fermions

$$c_l = \frac{\eta_{l+1} - i\eta_l}{\sqrt{2}}, \qquad l = 2, 4\ldots N-1. \quad (70)$$

To deal with the remaining $\eta^1$ we are obliged to introduce an additional "ghost" fermion field $\Phi$ such that if $\eta^1 = \frac{c_0 + c_0^\dagger}{\sqrt{2}}$, $\Phi = -\frac{c_0 - c_0^\dagger}{\sqrt{2}i}$. With this representation, for $N$ odd:

$$\hat{P} = 2^{\frac{N+1}{2}} \Phi \eta^1 \ldots \eta^N. \tag{71}$$

Since $\Phi$ commutes with the Hamiltonian, this also implies that for $N$ odd, the fermion operator

$$\mathcal{Q} = (-2i)^{N/2} \eta^1 \ldots \eta^N \tag{72}$$

commutes with Hamiltonian. This can be also verified directly. Thus, for odd $N$, the $Z_2$ invariance manifests itself as a supersymmetry

$$Z_2 \quad \Longleftrightarrow \quad [\hat{\mathcal{Q}}, H] = 0. \tag{73}$$

We now formulate the partition function as a functional integral

$$Z = \int \mathcal{D}\Psi \mathcal{D}\eta \, e^{-S}, \tag{74}$$

$$S = \int_0^\beta \left[ \frac{1}{2} (\eta \partial_\tau \eta + \sum_n \vec{\Psi}(n) \cdot \partial_\tau \vec{\Psi}(n)) + H \right] d\tau. \tag{75}$$

To proceed, we make a Hubbard Stratonovich transformation on the interaction term

$$-\frac{\tilde{J}}{2N} \left( i\vec{\eta} \cdot \vec{\Psi} \right)^2 \rightarrow \left[ iV(\tau)(\vec{\eta} \cdot \vec{\Psi}) + \frac{NV^2}{2\tilde{J}} \right], \tag{76}$$

where the $V(\tau)$ is a fluctuating real field. In the large $N$ limit, fluctuations of this field are small, and at $N = \infty$ the thermodynamics is determined by the mean-field saddle point $V = \frac{\tilde{J}}{N} \langle i\vec{\eta} \cdot \vec{\Psi}(0) \rangle$. This saddle point solution breaks the local $Z_2$ invariance, and at finite $N$ fluctuations in the sign of $V(\tau)$ play an important role in restoring the local symmetry.

The mean field Hamiltonian

$$H_{MF}(V) = \sum_k \epsilon_k \vec{\Psi}_{-k} \vec{\Psi}_k + V[i\vec{\eta} \cdot \vec{\Psi}(0)] + \frac{NV^2}{2J}, \tag{77}$$

$$\vec{\Psi}_k = N^{-1/2} \sum_j \vec{\Psi}(j) e^{-ikx_j},$$

describes an $N$-fold degenerate band hybridizing with a resonant level. At this level of approximation, the properties of the Fermi system are those of non-interacting resonant level of width $\Delta = \pi \rho V^2$. Using the resonant level free energy obtained in (55), we find

$$F_{MF} = -T \, \text{Tr} e^{-\beta H_{MF}} = N \left( F[\Delta] + \frac{\Delta}{2\pi \tilde{J} \rho} \right). \tag{78}$$

Minimizing this with respect to $\Delta$, we obtain

$$\psi(\frac{1}{2} + \frac{\Delta \beta}{2\pi}) - \ln \frac{T_K \beta}{2\pi} = 0, \tag{79}$$

where

$$T_K = D \exp \left[ -\frac{1}{\tilde{J}\rho} \right] \tag{80}$$

is the "Kondo temperature" which sets the characteristic energy scale of the mean-field properties. At low temperatures, $\Delta \longrightarrow T_K$.

Let us now discuss the fluctuations about the mean field theory. In the functional integral

$$Z = \int \mathcal{D}\Psi \mathcal{D}\eta \mathcal{D}V e^{-S}, \tag{81}$$

where

$$S = \int_0^\beta \left[ \frac{1}{2} (\eta \partial_\tau \eta + \sum_n \Psi(n) \partial_\tau \Psi(n)) + H_{MF}(V) \right] d\tau, \tag{82}$$

we may identify two classes of fluctuations: (i) small Gaussian fluctuations around the saddle point solution and (ii) rare tunneling events when the sign of the real field $V$ reverses ("kinks") as illustrated in Fig. 8.

In the large $N$ limit, the action associated with an order parameter "kink" scales as $N$, and hence the frequency of these fluctuations scales as $\Gamma(N) \propto \exp(-N)$. The small amplitude Gaussian fluctuations of frequency $\omega \gtrsim \Gamma$ will be unaffected by the rare large amplitude kinks, thus in the large $N$ limit we may consider the contributions to the functional integral from these fluctuations independently.

Consider first the effect of the Gaussian fluctuations. Expanding the action around the mean field theory we find

$$S = S_{MFT} + \frac{N}{2} \sum_{\nu_n} \Pi(\nu_n) \delta V(i\nu_n) \delta V(-\nu_n) + O(\delta V^3), \tag{83}$$

where $\Pi(\nu_n)$ is given by the fermion bubble

$$\Pi(i\nu_n) = \frac{1}{N} \langle (i\vec{\eta} \cdot \vec{\Psi})_{-\nu_n} (i\vec{\eta} \cdot \vec{\Psi})_{\nu_n} \rangle. \tag{84}$$

Since $T_K$ is the only energy scale in the problem, at low temperatures $\Pi(\nu_n)$ has the functional form

$$\Pi(\nu_n) = \rho f(|\nu_n|/T_K). \tag{85}$$

Though we can calculate $f(x)$, its detailed form is not of great importance. Since there is no continuous broken symmetry associated with the mean-field theory, these Gaussian fluctuations are gap-full at all frequencies and $f(x)$ is a positive analytic function of $x$. Gaussian fluctuations therefore generate small $1/N$ corrections to the dynamics that are analytic functions of the frequency, so they do not give rise to non-Fermi liquid behavior.

Next we consider the effect of the large discrete fluctuations in the sign of $V$. Let us divide the trajectory $V(\tau)$ into periods of time when $V = \pm V_0$ is almost constant, separated by brief periods when $V(\tau)$ is changing sign. The time scale of the sign change can be estimated by the rate at which $V(\tau)$ approaches the stationary value $\pm V_0$ after the transition has occurred. This rate can be estimated from (84): $1/\tau_0 \sim T_K$.

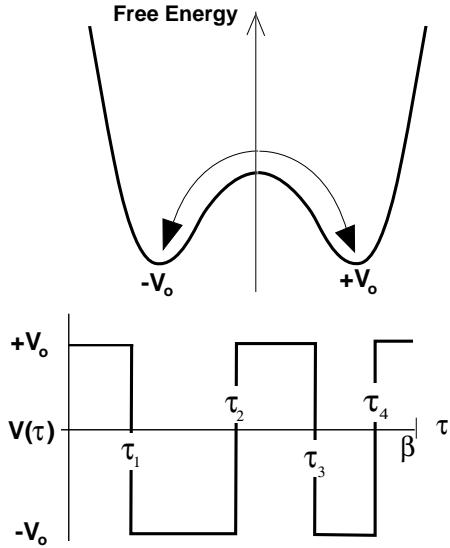

Fig. 8.: **Kink configuration.** Configuration with four kinks in the order parameter $V$ at a given site.

We should like to compute the action associated with $2n$ widely spaced sign changes in $V(\tau)$. We anticipate that there will be fixed amount of action associated with each transition, but there will be also Berry phase contributions that lead to long time interactions between kinks.

To calculate the overall Berry phase contributions to a tunneling trajectory we may use the "sudden approximation" in which each tunneling is considered to be instantaneous. The contribution to the path integral from a given tunneling trajectory of $2n$ kinks is then

$$\Gamma^{2n} Tr \left[ Te^{-\hat{S}} \right], \quad (86)$$

where $\Gamma$ is the amplitude of a single tunneling event and

$$\hat{S} = \int_{\tau_{2n}}^{\beta} H(V_0) d\tau - \int_{\tau_{2n-1}}^{\tau_{2n}} H(-V_0) d\tau \ldots - \int_{0}^{\tau_1} H(V_0) d\tau. \quad (87)$$

If we take advantage of the identity

$$Te^{-\int_{\tau_{2l-1}}^{\tau_{2l}} H(-V_0) d\tau} = \hat{P} Te^{-\int_{\tau_{2l-1}}^{\tau_{2l}} H(V_0) d\tau} \hat{P}, \quad (88)$$

we may write the contribution to the path integral from a given tunneling trajectory in the form

$$A_m[\{\tau_i\}] = \frac{\Gamma^m}{2} Tr \left[ T \prod_{j=1}^{m} \hat{P}(\tau_j) \sigma_x e^{-\int_0^{\beta} H(V_0) d\tau} \right], \quad (89)$$

where the trace includes the trace over a pseudo-spin variable $\sigma = \text{sgn}[V]$ that keeps track of the sign of $V_0$, and the normalization factor of $\frac{1}{2}$ takes care of the overcompleteness introduced by introducing both the Majorana spin representation and the additional ghost field $\Phi$.

The operator $\sigma_x$ converts pseudo-spin up to pseudo-spin down and vice versa, so $A_m[\{\tau_i\}]$ vanishes for $m$ odd, and includes the contributions from the two trajectories with flips at times $\{\tau_i\}$, starting out with either $V(0) = V_o$ or $V(0) = -V_o$. The total contribution to the functional integral from all such trajectories is then given by

$$A_n = \frac{1}{n!} \int d\tau_1 \ldots d\tau_n A_n[\{\tau_i\}]. \quad (90)$$

We may now recognize the sum of all $A_n$ as the perturbation expansion for the following time-ordered exponential

$$\sum_{n=0}^{\infty} A_n = \frac{1}{2} Tr \left[ Te^{-\int_0^{\beta} [H(V_0) + \Gamma \sigma_x \hat{P}] d\tau} \right]. \quad (91)$$

To estimate $\Gamma$, we may set it equal to a product of the attempt rate ($\sim T_K$) and the success rate of the tunneling process, the latter which becomes exponentially small for large $N$. To find the success rate with exponential accuracy we need to find the optimal trajectory leading from one minima to another which minimizes the action. We shall not do the explicit calculations here, instead we estimate it as an action associated with a very fast change of the sign. The exponential of this action is the overlap between the wave functions of the fermions in the ground states before and after the sign change, so

$$\Gamma \sim T_K \langle +|-\rangle. \quad (92)$$

We estimate $\langle +|-\rangle \sim \exp(-N/2)$.

We may rotate in the pseudo-spin space to transform $\sigma_x \to \sigma_z$. Since $\sigma_z$ commutes with the Hamiltonian, we may choose one sign of the pseudospin. The factor of 2 in the partition function that this gives rise to, cancels the normalization in (91). The effective Hamiltonian for low energy fluctuations then becomes

$$H_N = H(V_0) + \Gamma \hat{P}. \quad (93)$$

For even $N$,

$$H_N = H(V_0) + \tilde{\Gamma} \eta^1 \ldots \eta^N, \quad (94)$$

describes a Fermi liquid, with a weak interaction amongst the hybridized Fermi fields $\eta$. However, for odd N, the effective interaction couples the zero energy Majorana fermion $\Phi$, which is un-hybridized, to a composite of $N$ hybridized Majoranas $\eta^1 \ldots \eta^N$:

$$H_N = H(V_0) + \tilde{\Gamma} \Phi(\eta^1 \ldots \eta^N), \quad (95)$$

This result is qualitatively similar to that obtained in the strong coupling analysis of Section III, but it is now seen to be a feature of all $O(N)$ Kondo models with odd $N$. By repeating the perturbation arguments of the previous section, we now see that all odd $N$, there is a decoupled Majorana zero mode with a $\frac{1}{2} \ln 2$ zero point entropy.

Although the effective Hamiltonian (95) has the same qualitative form for all odd $N$ as the effective Hamiltonian (25) of the strong coupling limit, the large effect of the Majorana zero mode on conduction electrons is specific to $N = 3$. Consider, for example, the electron self-energy produced by weak scattering off the induced on-site interaction. The corresponding diagram is shown in Fig. 9. Since, the self energy scales as $1/\tau^{N-1}$, this implies

$$Im\Sigma(\omega) \sim \omega^{N-2}. \quad (96)$$

Marginal behavior only occurs for $N = 3$. The correction to the free energy for odd $N$

$$F \sim \int_0^\beta \frac{\Sigma(\tau)}{\tau} d\tau \sim \begin{cases} T^2 \ln T, & N = 3 \\ T^{N-1}, & N > 3. \end{cases} \quad (97)$$

Thus non-Fermi liquid behavior only develops for the physical $O(3)$ Kondo model and the unusual fermionic zero mode produces only weak corrections to Fermi liquid behavior at $N > 3$.

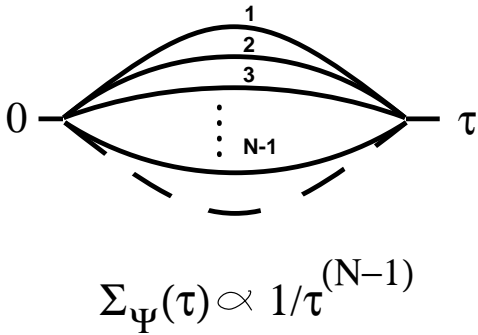

Fig. 9.: **Conduction electron self-energy in $O(N)$ Kondo impurity.** $\Sigma_\Psi(\tau)$ scales as $\frac{1}{\tau^{N-2}}$, which implies $Im\Sigma_\Psi(\omega) \propto \omega^{N-2}$ is marginally relevant when $N = 3$.

## V. THE O(N) LATTICE.

In this section we apply the approach developed in Sections III and IV to a lattice generalization of the two-channel Kondo model (2). The two complimentary methods of the last two sections have demonstrated how the driving force for non-Fermi liquid behavior of the two-channel impurity model is the formation of a fermionic zero mode. How specific is this behavior to impurity models, and do we expect similar type of behavior to persist in lattice generalizations of these models?

For more than one two-channel Kondo impurity, backscattering between spin-sites will couple "right" and "left" moving currents, so the kinematic spin-charge decoupling present in the original model is lost. In a two channel Kondo lattice, the main effect of the lattice will thus be to cut-off the non-Fermi liquid behavior by eliminating spin-charge decoupling.

In this section we consider the effect of the lattice on the zero-mode behavior of the $O(3)$ Kondo model. Unlike the two-channel Kondo lattice, here the complications of these spin-charge coupling effects do not arise. We begin by writing down the "$O(3)$" Kondo lattice model

$$H = -t \sum_{j,a_l} \left[ c^\dagger{}_j c_{j+a_l} + \text{H.c} \right] + J \sum_j \vec{S}_j \cdot [\vec{\sigma}_j + \vec{\tau}_j], \quad (98)$$

where $j$ represents the site on a three dimensional cubic lattice,

$$\vec{\sigma}_j = c^\dagger{}_j \vec{\sigma} c_j,$$
$$\vec{\tau}_j = \tilde{c}^\dagger{}_j \vec{\tau} \tilde{c}_j, \quad (99)$$

denote the spin and isospin density at site $j$ and $a_l$ ($l = 1, 2, 3$ denotes the three unit lattice vectors. By using the Majorana decoupling of previous sections:

$$c_j = \frac{e^{i\vec{Q}\cdot\vec{x}_j}}{\sqrt{2}} \left( \Psi_j^0 + i\vec{\Psi}_j \cdot \vec{\sigma} \right) \begin{pmatrix} 0 \\ -i \end{pmatrix}, \quad (100)$$

where $\vec{Q} = (\pi/2, \pi/2, \pi/2)$, this model can be re-written

$$H = -it \sum_{j,a_l} [\vec{\psi}_j \cdot \vec{\psi}_{j+a_l}] - iJ \sum_j \vec{S}_j \cdot [\vec{\psi}_j \times \vec{\psi}_j] + H_o, \quad (101)$$

where the decoupled scalar degrees of freedom, described by

$$H_o = -it \sum_{j,a_l} \psi_j^{(0)} \psi_{j+a_l}^{(0)}, \quad (102)$$

may be neglected. By writing the local moments in a Majorana representation, $\vec{S}_j = -\frac{i}{2}\vec{\eta}_j \times \vec{\eta}_j$, this model may be cast in a manifestly $O(3)$ symmetric form

$$H = -it \sum_{j,a_l} \left[ \vec{\psi}_j \cdot \vec{\psi}_{j+a_l} \right] - J \sum_j [i\vec{\eta}_j \cdot \vec{\psi}_j]^2. \quad (103)$$

Actually, both forms of this model are of interest. The model written in form (98) involves a coupling between pair, and spin degrees of freedom. Such a model appears very naturally as the low-energy model for an odd-frequency[28] superconductor with composite off-diagonal order.[29] In the more symmetric form, (103), the model can be compactly generalized to an $O(N)$ model, by extending the number of components of the vector fermions from 3 to $N$, as was done in the last section:

$$H = -it \sum_{j,a_l} [\vec{\psi}_j \cdot \vec{\psi}_{j+a_l}] - \frac{\tilde{J}}{2N} \sum_j [i\vec{\eta}_j \cdot \vec{\psi}_j]^2. \quad (104)$$

By examining this model around the large $N$ limit, following the methods of the last section, we can learn how the fermionic zero-mode and the non-Fermi liquid behavior are modified in a lattice environment.

Following the procedure introduced in Section IV we decouple the exchange interaction by the Hubbard-Stratonovich field $V(\vec{x},\tau)$ to obtain the following action:

$$S = \int d\tau \sum_{\vec{x}} \left[\frac{1}{2}\left(\vec{\eta}_j \partial_\tau \vec{\eta}_j + \vec{\psi}_j \partial_\tau \vec{\psi}_j\right) + H_j\right],$$
$$H_j = it\sum_{a_l}[\vec{\psi}_j \cdot \vec{\psi}_{j+a_l}] + iV_j(\vec{\eta}_j \cdot \vec{\psi}_j) + \frac{NV_j^2}{J}.$$

In the mean-field approximation the $V$-field acquires a non-zero average $V_o = \langle V \rangle$. To find we need to solve the equation

$$\frac{\delta F(V_o)}{\delta V_o} = \frac{NV_o}{J} + i\langle(\vec{\eta}_j \cdot \vec{\psi}_j)\rangle = 0, \quad (105)$$

where the fermionic Green's function is calculated against the background of a fixed uniform $V$. The mean-field Green's function on this background

$$\hat{G}(\omega_n, \vec{k}) = \begin{pmatrix} \langle\langle\psi^a\psi^a\rangle\rangle & \langle\langle\psi^a\eta^a\rangle\rangle \\ \langle\langle\eta^a\psi^a\rangle\rangle & \langle\langle\eta^a\eta^a\rangle\rangle \end{pmatrix}, \quad (106)$$

is given by

$$\hat{G}(\omega_n, \vec{k}) = \begin{pmatrix} i\omega_n - \epsilon_{\vec{k}} & iV \\ -iV & i\omega_n \end{pmatrix}^{-1}, \quad (107)$$

where $\epsilon_{\vec{k}} = 2t\sum_{l=x,y,z} \sin p_l$. The quasiparticle spectrum, determined from the eigenvalue equation $\det[\hat{G}(E(\vec{k}),\vec{k})^{-1}] = 0$, is

$$E_{\vec{k}\lambda} = \frac{\epsilon_{\vec{k}}}{2} + \lambda\sqrt{\epsilon^2(\vec{k})/4 + V_o^2}, \quad (\lambda = \pm 1) \quad (108)$$

contains a small gap:

$$\Delta_g = \frac{V_o^2}{|\epsilon(\vec{Q})|}. \quad (109)$$

The mean-field free energy per unit cell is then

$$F_{MF}[V_o] = -\frac{NT}{2}\sum_{\vec{k},\lambda}\ln\left[2\cosh(\beta E_{\vec{k}\lambda}/2)\right]$$
$$+ \frac{NV_o^2}{2J}. \quad (110)$$

Minimizing with respect to $V_o$ yields

$$V_o = J\sum_{\vec{k}} \tanh(\beta E_{\vec{k}(+)}/2)\frac{\lambda V_o}{\sqrt{\epsilon_{\vec{k}}^2/4 + V_o^2}}. \quad (111)$$

Replacing the momentum integral by an energy integral near the Fermi energy $\sum_{\vec{k}} \rightarrow \rho\int d\epsilon$ and performing the integral at $T=0$ yields

$$V_o = 2\rho J V_o \ln\left(\frac{D}{|V_o|}\right), \quad (112)$$

where $D \approx 6t$ is the upper cut-off and $\rho$ is the density of states at the Fermi level of itinerant electrons. Solving this equation for $V_o$ with logarithmic accuracy we find

$$V_o \approx De^{-\frac{1}{2\rho J}} = \sqrt{T_K D} \quad (113)$$

where $T_K = D^{-\frac{1}{\rho J}}$, so that the gap $\Delta_g = T_K$.

Clearly, the $Z_2$ symmetry implies that the mean-field free energy is independent of the sign of $V_j$ at each site, and beyond mean-field theory, $V_j(\tau)$ will fluctuate between $V_o$ and $-V_o$. We can apply the technique developed in Section IV to take into account the effect of these sign fluctuations. In particular, it was shown that this effect can be emulated by the integration over an additional Majorana ghost field $\Phi$:

$$\int DV e^{-i\int d\tau V(\vec{\eta}\cdot\vec{\psi})}$$
$$\propto \int D\Phi \exp\left[-\int_0^\beta d\tau[\frac{1}{2}\Phi\partial_\tau\Phi + +g\Phi\eta^1\eta^2\eta^3]\right], \quad (114)$$

where the coupling constant $g$ is physically identified with the inverse tunneling rate for a local change of sign in the $V(n,\tau)$ field. As local processes are controlled by the energy scale $T_K$, we may estimate $g \sim T_K$. The scale $T_K$ is exponential in the bare coupling constant $J$, so the result (114) is essentially non-perturbative.

The fermionic variable $\Phi$ represents low-lying excitations of the theory. In order to derive the effective action for these excitations, we need to integrate out high-energy degrees of freedom described by the gapfull modes $\eta_i, \psi_i$ ($i=1,2,3$). The mode $\Phi$ interacts with these gapfull modes via

$$H_{int} = g\sum_j \left[\Phi_j \eta_j^1 \eta_j^2 \eta_j^3\right]. \quad (115)$$

The simplest diagram for the self-energy part of $\Phi$ is shown in Fig.(10). It results in the dispersion

$$\Sigma(k) \sim \left(\sum_{l=1}^3 \sin k_l\right)\frac{g^2}{T_K}. \quad (116)$$

Thus the low energy properties of the two-channel Kondo lattice are determined by a single band of real fermions with a bandwidth of order of the Kondo temperature. Since the fermion is real, the total amount of entropy accumulated in this band is $\ln\sqrt{2}$ per site, coinciding with the residual entropy of the single impurity problem. The effective inter-site interaction induced by virtual three-body fluctuations removes the degeneracy of the single impurity problem and results in the coherent dispersion of the $\Phi$-mode.

At low temperatures the propagating $\Phi$-mode now leads to a linear specific heat. Since the matrix elements of spin operators and $\Phi$-states are zero, there are no low-energy spin-excitations. Remarkably, the two-channel Kondo lattice represents a spin paramagnet with a finite spectral gap $2T_K$. Above the energy gap a single spin kink is a composite excitation consisting of two gapfull Majorana fermions. In this respect this model is similar to the spin-1/2 Heisenberg chain.

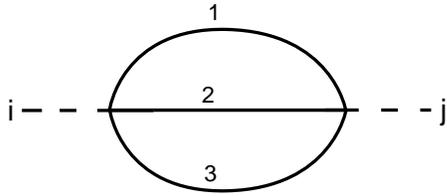

Fig. 10.: **Inter-site coupling of three body bound-state.** The three-body bound-state can disperse from site to site via virtual fluctuations into the conduction sea.

In the weak coupling limit of this model, the three-body bound-states develop coherence at a temperature $g^2/T_K \sim T_K$. Clearly then, the simplest lattice generalization of the $O(3)$ model will not display any non-Fermi liquid like behavior. In the strong-coupling limit however, there is a possibility of a large separation between the the scale $g^2/T_K$ and $T_K$, leading to a regime where the three-body bound-states are localized and conduction electron self-energies are marginal. This is an interesting issue that we return to in the final section.

## VI. CONCLUSION

A goal of this paper was to explore the non-Fermi liquid properties of the two-channel Kondo model by using a "stripped-down" version of the this model that describes the decoupled spin degrees of freedom. The compactified model that arose from these considerations provides a new perspective on the two-channel Kondo model and opens up a new family of $O(N)$ Kondo models which generalize the essential physics beyond the realm of the original model. In this final section we discuss further applications of our methods and touch upon some of the interesting questions that arise in connection with the $O(N)$ Kondo lattice model.

The key observation of this paper was the equivalence of the two-channel impurity Kondo model

$$H' = i\,t \sum_n [a^\dagger_{\lambda\sigma}(n+1)a_{\lambda\sigma}(n) - \mathrm{H.c.}]$$
$$+ J[\vec{\sigma}_1(0) + \vec{\sigma}_2(0)] \cdot \vec{S}_d, \qquad (117)$$

with its compactified counter-part

$$H = i\,t \sum_n [c^\dagger_\sigma(n+1)c_\sigma(n) - \mathrm{H.c.}]$$
$$+ J[\vec{\sigma}(0) + \vec{\tau}(0)] \cdot \vec{S}_d, \qquad (118)$$

under the mapping

$$\begin{aligned}\vec{\sigma}_1(x) &\longleftrightarrow \vec{\sigma}(x),\\ \vec{\sigma}_2(x) &\longleftrightarrow \vec{\tau}(x),\end{aligned} \qquad (119)$$

that resulted from spin-charge decoupling. The universal spin physics of both models is the same, but by preserving spin-charge decoupling in the compactified model, the geometry of the scaling flows away from weak coupling was deformed, moving the location of the NFL fixed point from intermediate coupling in the old coupling constant, to infinite-coupling of the redefined coupling constant.

Several other applications of this procedure have yet to be explored. One clear application is the use of the the new formulation in an actual numerical Wilson renormalization calculation. Wilson's original numerical renormalization group approach to the one-channel Kondo model took great advantage of a strong-coupling expansion to connect the numerical results to the strong-coupling fixed point. This can not be done in the standard formulation of the two-channel Kondo model, but it can now be done with the new formulation. This then opens the possibility for a direct numerical calculation of the marginal Fermi liquid spin-susceptibility of this model.

Another simple application of this reformulation is to the channel-anisotropic two channel Kondo model, where the original interaction

$$H_{int} = \left[J_1\vec{\sigma}_1(0) + J_2\vec{\sigma}_2(0)\right] \cdot S_d, \qquad (120)$$

may be compactified into

$$H_{int} = \left[J_1\vec{\sigma}(0) + J_2\vec{\tau}(0)\right] \cdot S_d. \qquad (121)$$

Bethe-ansatz work on this model suggests that the character of the paramagnetic fixed point changes subtly in response to channel anisotropy.[31] Unlike the original model, the strong-coupling limit of the compactified channel-anisotropic model is a stable fixed point. The "Fermi liquid" of spins that results from such a description appears to very naturally described as a band of vector and scalar Majorana fermions.[32]

Perhaps the most fascinating aspect of the compactified model concerns the appearance of three-body bound-states and their close link with marginal Fermi liquid behavior. The original marginal Fermi liquid hypothesis was advanced as a phenomenological framework to describe the normal state of cuprate superconductors.[2] An essential feature of the marginal Fermi liquid ansatz is the appearance of a two particle susceptibility

$$\chi''(\omega) \propto \begin{cases} \frac{\omega}{T}, & (\omega << T) \\ \mathrm{constant}, & (\omega >> T) \end{cases} \qquad (122)$$

and a related electron self-energy

$$\Sigma(\omega) \sim \omega \ln(\max[\omega, T]). \qquad (123)$$

It proves remarkably difficult to construct even a toy model that furnishes a marginal self-energy down to arbitrarily low energies. A scale-invariant self-energy of this form requires a massless excitation. Almost inevitably, beyond leading orders in perturbation theory, a nominally massless excitation develops a mass. In an attempt to try to overcome this difficulty, Varma and Ruckenstein[33] tentatively suggested that the massless excitation responsible for marginal behavior might have its origins in a three-fermion bound state. Quite remarkably, the marginal Fermi liquid behavior associated with the $O(3)$ Kondo model is directly related to the formation of a three-body bound state. The fixed point Hamiltonian

$$H^* = it \sum_{n=1}^{n=\infty} \vec{\Psi}(n+1) \cdot \vec{\Psi}(n)$$
$$+ \alpha \Phi \big[ \Psi^{(1)}(1) \Psi^{(2)}(1) \Psi^{(3)}(1) \big], \quad (124)$$

owes its marginal Fermi liquid properties to the coupling of a Majorana three-body bound-state to the continuum. The *real* character of the three-body operator $\Phi = \Phi^\dagger$ means that its square is a c-number $\Phi^2 = \frac{1}{2}\{\Phi, \Phi^\dagger\}$, so that any "mass term" of the form $m^2 \Phi^2 = m^2/2$ can be absorbed into the Ground-state energy, which prevents the development of a mass term associated with this excitation. A Majorana three-body bound-state in an impurity model is thus massless in to all orders in perturbation theory. This essential Majorana character of the three-body bound state thus ensures that:

- The response function of the three-body bound-state retains a zero-energy pole, giving rise completely scale-invariant response functions.

- The joint susceptibility $\chi_{e\Phi}(\omega)$ of the electrons and the three-body bound-state is marginal, with no characteristic energy scale

$$\chi_{e\Phi}(\omega)'' = \rho \left[ \frac{1}{2} \tanh\left(\frac{\omega}{T}\right) \right],$$
$$\propto \begin{cases} \frac{\omega}{T} & (\omega \ll T) \\ \text{constant} & (\omega \gg T) \end{cases} \quad (125)$$

The generalization of the strong-coupling Hamiltonian (124) to a lattice environment is particularly interesting:

$$H^* = it \sum_{\hat{j}, \hat{a}} \vec{\Psi}_{\hat{j}+\hat{a}} \cdot \vec{\Psi}_{\hat{j}}$$
$$+ \alpha \sum_j \Phi(j) \Psi^{(1)}(j) \Psi^{(2)}(j) \Psi^{(3)}(j). \quad (126)$$

This is a model with "pre-formed" three-body bound-states. Clearly, marginal Fermi liquid behavior will persist in this model down to temperatures $T \sim T^*$, where the three-body bound-state begins to propagate coherently. This temperature $T^*$ is set by the size of the coupling constant

$$T^* = \frac{\alpha^2}{t}, \quad (127)$$

providing a rather interesting toy model for detailed calculations of marginal Fermi liquid behavior in a lattice environment.

Another suggestive aspect of this work this work concerns its possible links with odd-frequency pairing.[28,29] The semi-classical methods that become exact in the large-$N$ limit of the $O(N)$ Kondo model are almost identical to those used to develop a mean-field theory for odd-frequency pairing in the Kondo lattice.[30] Consider the family of Kondo lattice models

$$H[\lambda] = \sum_{\vec{k}} \epsilon_{\vec{k}} \psi^\dagger_{\vec{k}\sigma} \psi_{\vec{k}\sigma} + J \sum_j [\vec{\sigma}(j) + \lambda \vec{\tau}(j)] \cdot \vec{S}_d[j], \quad (128)$$

where $\vec{\sigma}(j)$ and $\vec{\tau}(j)$ denote the spin and isospin density at site $j$ respectively. This type of model was recently suggested as a low-energy Hamiltonian for an odd-frequency superconductor,[29] where the $\lambda$ dependent term develops in response to the growth of composite order $\langle \sigma_a(j) \tau_b(j) \rangle = R_{ab}(j)$ between spin and charge degrees of freedom. The lambda term can be considered analogous to a Weiss field for odd-frequency pairing. The model described by $H[1]$ is the $O(3)$ Kondo lattice model considered in this article; $H[0]$ is the conventional Kondo lattice model. Remarkably, if we use a Majorana representation of the local moments, $\vec{S}_d = -\frac{i}{2} \vec{\eta}_j \times \vec{\eta}_j$ and the conduction electrons, then the same mean-field ansatz

$$-iJ\langle \vec{\eta}_j \cdot \vec{\psi}_j \rangle = V, \quad (129)$$

can be used to extend the semi-classical methods used in this article to all values of $\lambda$. The mean-field theory contains a spin and charge gap for each value of $\lambda$, and as $\lambda$ evolves, the state evolves continuously. For $\lambda = 1$, this semi-classical theory corresponds to that described in this article. In the limit $\lambda \to 0$, the mean-field theory provided here corresponds to a broken-symmetry state with composite order and odd-frequency pairing.[30] These results suggest that there may be a whole range of qualitatively similar physics that evolves continuously from the $O(3)$ Kondo lattice model to the conventional Kondo lattice model.

The authors are particularly grateful to Natan Andrei and Eduardo Miranda for detailed discussions related to this work. Discussions with Elihu Abrahams, Antoine Georges and Anirvan Sengupta also helped the evolution of our ideas. This work was supported by NSF grant DMR-93-12138, travel funds from NATO grant CRG 940040 and the EPSRC, UK.

## APPENDIX A: Validity of the compactification procedure.

Here, we discuss in more detail, the compactified formulation of the two-channel Kondo model. We begin by linearizing the conduction bands around the Fermi energy to write the two channel Kondo model in the relativistic form

$$H = \sum_{\lambda=1}^{M} \left[ -iv_F \int dx \psi^\dagger{}_\lambda \nabla \psi_\lambda + J\vec{S}_d \cdot \psi^\dagger{}_\lambda(0)\vec{\sigma}\psi_\lambda(0) \right]. \quad (130)$$

This model describes $M$ chains of right-moving electrons with group velocity $\lambda$, each interacting with a single spin $\vec{S}_d$ at the origin. Each right moving electron spinor $\psi_\lambda(x)$ represents an electron scattering off a Kondo impurity in an s-wave scattering channel. The properties of this model are determined by the correlation functions of the total conduction electron spin density in the non-interacting system. For instance, consider the perturbation expansion of the partition function

$$\frac{Z}{Z_o} = 1 + \sum_{n=1}^{\infty} \frac{J^n}{n!} \mathcal{B}_n. \quad (131)$$

Here

$$\mathcal{B}_n = \int_0^\beta \prod_{j=1}^n d\tau_j \langle T\sigma_T^{a_1}(\tau_1)\ldots\sigma_T^{a_n}(\tau_n)\rangle_0$$
$$\times \langle TS_d^{a_1}(\tau_1)\ldots S_d^{a_n}(\tau_n)\rangle_0, \quad (132)$$

where the subscript (0) denotes correlation functions in the non-interaction system and

$$\sigma_T^a(x) = \sum_{\lambda=1,M} \psi^\dagger{}_\lambda(x)\sigma^a \psi_\lambda(x), \quad (133)$$

is the total conduction electron spin density. In a one-dimensional problem, the spin densities behave as independent degrees of freedom; for the $M$ channel problem, the the Kinetic energy decouples into a sum of independent terms $H_{kin} = \sum_{\lambda=1,M} H_\lambda + \ldots$, where $H_\lambda = v_F \sum_q : \vec{\sigma}_\lambda(q) \cdot \vec{\sigma}_\lambda(-q) :$ is a bilinear in the Fourier transform of the spin density $\vec{\sigma}_\lambda(q) = \sum_k \psi^\dagger{}_{\lambda k}\vec{\sigma}\psi_{\lambda k+q}$ at momentum $q$, where $\psi_{\lambda k}$ denotes an electron of momentum $k$. The dynamics of these modes are set exclusively by their commutation algebras ("current algebra"). The spin density satisfies the current algebra

$$[\sigma_\lambda^a(q), \sigma_{\lambda'}^b(-q')] = \left[ 2i\epsilon_{abc}\sigma_\lambda^c(q-q') + q\delta^{ab}\delta_{q,q'} \right]\delta_{\lambda\lambda'} \quad (134)$$

For the case where $M = 2$, the spin dynamics of two spin channels may be mapped onto the spin and isospin dynamics of a *single* chain. Consider the single chain model where both spin and isospin interact with a local moment

$$H = -iv_F \int dx \psi^\dagger{}_s \nabla \psi_s dx + J\vec{S}_d \cdot \left[ \vec{\sigma}(0) + \vec{\tau}(0) \right], \quad (135)$$

where

$$\vec{\sigma}(x) = c_\alpha^\dagger(x)\vec{\sigma}_{\alpha\beta}c_\beta(x)$$

is the spin density at $x$,

$$\vec{\tau}(x) = \tilde{c}_\alpha^\dagger(x)\vec{\sigma}_{\alpha\beta}\tilde{c}_\beta(x)$$

is the isospin density at $x$, written in terms of the Nambu spinor

$$\tilde{\psi}(x) = \begin{pmatrix} \psi_\uparrow(x) \\ \psi_\downarrow^\dagger(x) \end{pmatrix}. \quad (136)$$

For one-dimensional electrons moving with linear dispersion, spin and isospin form two independent current algebras

$$[\sigma^a(q), \sigma^b(-q')] = 2i\epsilon_{abc}\sigma^c(q-q') + q\delta^{ab}\delta_{q,q'} , \quad (137)$$

$$[\tau^a(q), \tau^b(-q')] = 2i\epsilon_{abc}\tau^c(q-q') + q\delta^{ab}\delta_{q,q'} , \quad (138)$$

$$[\sigma^a(q), \tau^b(q')] = 0, \quad (139)$$

where the label $q$ denotes the Fourier transform of the density at wave vector $q$. Since the electron Kinetic energy can be expressed as the sum of two independent bilinears of charge and spin current densities,

$$-iv_F \int (\psi^\dagger{}_s \nabla \psi_s) dx$$
$$= v_F \sum_q : \vec{\sigma}(q) \cdot \vec{\sigma}(-q) + \vec{\tau}(q) \cdot \vec{\tau}(-q) :, \quad (140)$$

it follows that spin and isospin behave as two completely independent spin degrees of freedom with *precisely* the same dynamics as the spin densities of two independent chains. When interactions are introduced that couple to to $\sigma_1(x)$ and $\sigma_2(x)$, their effect will be identical in a compactified model where we have made the replacement

$$\begin{aligned}\vec{\sigma}_1(x) &\rightarrow \vec{\sigma}(x), \\ \vec{\sigma}_2(x) &\rightarrow \vec{\tau}(x).\end{aligned} \quad (141)$$

For example, in the case of channel symmetry in the compactified model, the local moment couples to the total spin density $\vec{\mathcal{T}}(x) = \vec{\sigma}(x) + \vec{\tau}(x)$. This operator obeys same current algebra as $\vec{\sigma}_T(x) = \vec{\sigma}_1(x) + \vec{\sigma}_2(x)$ in the original two channel model, so both operators have precisely the same correlation functions in the absence of interactions. Thus in the expansion of the partition function of the compactified two channel model

$$\frac{Z}{Z_0} = 1 + \sum_{n=1}^{\infty} \frac{J^n}{n!}\mathcal{C}_n, \quad (142)$$

where

$$\mathcal{C}_n = \int_0^\beta \prod_{j=1}^n d\tau_j \langle T\mathcal{T}_T^{a_1}(\tau_1)\ldots\mathcal{T}_T^{a_n}(\tau_n)\rangle_0$$
$$\times \langle TS_d^{a_1}(\tau_1)\ldots S_d^{a_n}(\tau_n)\rangle_0, \quad (143)$$

the correspondence of correlation functions guarantees that the expressions in (132) and (143) are equal: $\mathcal{C}_n = \mathcal{B}_n$, establishing the equivalence of the two partition functions. Similar perturbative arguments can be made for any response function involving spin degrees of freedom, even in the case of channel anisotropy.

Of course, if we now introduce a cut-off into these models by rewriting them as one-dimensional lattice models with band-width $2t$, the equivalence of the two channel model with its compactified counterpart is only guaranteed in the limit $J/t \ll 1$, at temperatures and frequencies $\ll t$. If we apply a Wilson renormalization procedure to the two models, then although the two calculations will predict the same low temperature properties, the details of renormalization flows will differ beyond weak coupling. General scaling principles ensure that the the universal low energy dynamics predicted by the Wilson procedure will match up for the two models.

## APPENDIX B: Simple manipulations of Majorana Fermions

Here we present some of the technical manipulations required to express the spin and isospin density at the origin in terms of the Majorana fermion operators. We begin with the expression for the Fermi field at the origin

$$c = \frac{1}{\sqrt{2}} \left( \Psi^0 + i\vec{\Psi} \cdot \vec{\sigma} \right) \begin{pmatrix} 0 \\ -i \end{pmatrix}, \quad (144)$$

or more explicitly

$$\begin{pmatrix} c_\uparrow \\ c_\downarrow \end{pmatrix} = \frac{1}{\sqrt{2}} \begin{pmatrix} \Psi^1 - i\Psi^2 \\ -\Psi^3 - i\Psi^0 \end{pmatrix}, \quad (145)$$

where $(\Psi^0, \vec{\Psi})$ are four real (or Majorana) components which satisfy

$$\Psi^a = (\Psi^a)^\dagger,$$
$$\{\Psi^a, \Psi^b\} = \delta^{ab}. \quad (146)$$

To deal simultaneously with the spin and isospin operators, it is convenient to rewrite the Fermi spinor as a Balian-Werthamer four spinor, as follows

$$\mathcal{C} = \begin{pmatrix} c_\uparrow \\ c_\downarrow \\ c^\dagger_\downarrow \\ -c^\dagger_\uparrow \end{pmatrix} = \frac{1}{\sqrt{2}} \left( \Psi^0 + i\vec{\Psi} \cdot \vec{\sigma} \right) \mathcal{Z}, \quad (147)$$

where

$$\mathcal{Z} = \begin{pmatrix} 0 \\ -i \\ i \\ 0 \end{pmatrix}. \quad (148)$$

In terms of this four-spinor, the spin and isospin operators are

$$\vec{\sigma} = \frac{1}{2}\mathcal{C}^\dagger \vec{\underline{\sigma}} \mathcal{C},$$
$$\vec{\tau} = \frac{1}{2}\mathcal{C}^\dagger \vec{\underline{\tau}} \mathcal{C}, \quad (149)$$

where

$$\vec{\underline{\sigma}} \equiv \vec{\sigma} \otimes 1 = \left[ \begin{array}{c|c} \vec{\sigma} & \\ \hline & \vec{\sigma} \end{array} \right], \quad (150)$$

and

$$(\underline{\tau}_1, \underline{\tau}_2, \underline{\tau}_3) = \left( \left[ \begin{array}{c|c} & 1 \\ \hline 1 & \end{array} \right], \left[ \begin{array}{c|c} & -i \\ \hline i & \end{array} \right], \left[ \begin{array}{c|c} 1 & \\ \hline & -1 \end{array} \right] \right). \quad (151)$$

When sandwiched between the spinors $\mathcal{Z}$ and $\mathcal{Z}^\dagger$, these matrices give the following expectation values:

$$\mathcal{Z}^\dagger \vec{\underline{\sigma}} \mathcal{Z} = \mathcal{Z}^\dagger \vec{\underline{\tau}} \mathcal{Z} = 0,$$
$$\mathcal{Z}^\dagger \underline{\sigma}^a \underline{\sigma}^b \mathcal{Z} = 2\delta^{ab},$$
$$\mathcal{Z}^\dagger \underline{\sigma}^a \underline{\tau}^b \mathcal{Z} = -2\delta^{ab},$$
$$\mathcal{Z}^\dagger \underline{\sigma}^a \underline{\sigma}^b \underline{\sigma}^c \mathcal{Z} = 2i\epsilon^{abc},$$
$$\mathcal{Z}^\dagger \underline{\sigma}^a \underline{\sigma}^b \underline{\tau}^c \mathcal{Z} = -2i\epsilon^{abc}. \quad (152)$$

With these results, we can expand the expressions in (149) to obtain

$$\sigma^a = \frac{1}{4}\mathcal{Z}^\dagger \left[ (\Psi_o - i\vec{\underline{\sigma}} \cdot \vec{\Psi}) \underline{\sigma}^a (\Psi_o + i\vec{\underline{\sigma}} \cdot \vec{\Psi}) \right] \mathcal{Z}$$
$$= i \left[ \Psi^0 \vec{\Psi} - \frac{1}{2}\vec{\Psi} \times \vec{\Psi} \right]^a, \quad (153)$$

and

$$\tau^a = \frac{1}{4}\mathcal{Z}^\dagger \left[ (\Psi_o - i\vec{\underline{\sigma}} \cdot \vec{\Psi}) \underline{\tau}^a (\Psi_o + i\vec{\underline{\sigma}} \cdot \vec{\Psi}) \right] \mathcal{Z}$$
$$= -i \left[ \Psi^0 \vec{\Psi} + \frac{1}{2}\vec{\Psi} \times \vec{\Psi} \right]^a. \quad (154)$$

Combining these two results, we find

$$\vec{\sigma} + \vec{\tau} = -i\vec{\Psi} \times \vec{\Psi}. \quad (155)$$